\documentclass{jkas}

\def\year{2018} % publication year
\def\month{June} % publication month
\def\volume{51} % journal volume
\def\issue{3} % journal issue
\def\beginpage{73} % first page of article
\def\endpage{88} % last page of article
\def\received{May 9, 2018} % date paper was received by JKAS
\def\accepted{June 30, 2018} % date of acceptance
\date{Received \received; accepted \accepted}

\def\lesssim{\mathrel{\hbox{\rlap{\hbox{\lower4pt\hbox{$\sim$}}}\hbox{$<$}}}}
\def\gtrsim{\mathrel{\hbox{\rlap{\hbox{\lower4pt\hbox{$\sim$}}}\hbox{$>$}}}}
\def\apj{ApJ}
\def\apjs{ApJS}
\def\aj{AJ}
\def\aap{A\&\hskip-1pt A}

\def\mnras{MNRAS}

\usepackage{flushend}
\title{Luminosity Profiles of Prominent Stellar Halos}
\author{Hong Bae Ann and Hyeong Wook Park}
\affil{Pusan National University, 2 Busandaehak-ro, Geumjeong-gu, Busan 46241, Korea; \email{hbann@pusan.ac.kr}}

\def\corrauthor{H. B. Ann}
\def\runningauthor{Ann \& Park}
\def\runningtitle{Luminosity Profiles of Prominent Stellar Halos}
\def\keywords{galaxies: general --- galaxies: photometry --- galaxies: structure
}

\def\abstracttext{
We present a sample of 54 disk galaxies which have well developed
extraplanar structures. We selected them using visual inspections from
the color images of the Sloan Digital Sky Survey. Since the sizes of the
extraplanar structures are comparable to the disks, they are considered
as prominent stellar halos rather than large bulges. A single S\'ersic profile
fitted to the surface brightness along the minor-axis of the disk
shows a luminosity excess in the central regions for the majority of
sample galaxies. This central excess is considered to be caused by the
central bulge component.
The mean S\'ersic index of the single component model is $1.1\pm0.9$.
A double S\'ersic profile model that employs $n=1$ for the inner region,
and varying $n$ for the outer region, provides a better fit than the single
S\'ersic profile model. For a small fraction of galaxies, a S\'ersic profile
fitted with $n=4$ for the inner region gives similar results. There is a weak
tendency of increasing $n$ with increasing luminosity and central velocity
dispersion, but there is no dependence on the local background density.
}

\begin{document}
\jkashead %% set title, authors, abstract, etc.

%%%%%%%%%%%%%%%%%%%%%%%%%%%%%%%%%%%%%%%%%%%%%%%%%%%%%%%%%%%%%%%%%%%%%
%%% BEGIN MAIN TEXT HERE %%%%%%%%%%%%%%%%%%%%%%%%%%%%%%%%%%%%%%%%%%%%
%%%%%%%%%%%%%%%%%%%%%%%%%%%%%%%%%%%%%%%%%%%%%%%%%%%%%%%%%%%%%%%%%%%%%

\section{Introduction} \label{sec:intro}

Stellar halos are thought to be ubiquitous in disk galaxies \citep{sac94,
mor97, leq98, abe99, zib04, zibfer04, zib04, mcc06, cha06, kal06, set06, deJ07, mou07, hel08, jab10, iba14}. In the framework of the $\Lambda$ cold dark
matter ($\Lambda$CDM) cosmology \citep{spr06}, the existence of stellar halos
around disk galaxies is very natural because they are byproducts of
hierarchical galaxy formation. They are thought to be formed by the
accretion and
disruption of dwarf satellites \citep{bul05, aba06, fon06, del08, fon08, joh08, gil09, coo10,zol10}. There is mounting evidence that the accretion and
disruption of infalling satellites play a major role in building the
stellar halos around disk galaxies \citep{sea78, joh96, hel99, fer02, bul05, iba07, joh08, rom16}.
On the other hand, gas rich mergers in the early phase of the galaxy
assembly \citep{bro04}, and $in-situ$ star formation \citep{fon11} were also
proposed as explanations of the origin of stellar halos.
The ejection of $in-situ$ stars formed at high redshift ($z \gtrsim 3$) by
subsequent major \citep{zol09} and minor mergers \citep{pur10} also
could contribute to stellar halos.

The surface brightness of the stellar halos of spiral galaxies is typically
10 magnitudes below the night sky background \citep{bak12}, making them
difficult to observe. In the case of the Milky way, its halo luminosity
is $\sim1$ per cent of the total luminosity, and other spirals usually have halo luminosities similar to the Milky Way \citep{cou11}.
Since the surface brightness level needed to analyze the structure of stellar
halos is so faint ($\sim30$ mag arcsec$^{-2}$), it requires highly precise
observations and data reductions are required to
prevent observational artifacts.
The first detection of the stellar halo around disk galaxies was reported for
NGC5907 by \citet{sac94}, and confirmed by \citet{leq96} and \citet{leq98}.
\citet{zib04} applied an image stacking technique to disk galaxies
to derive the structural parameters of their stellar halos.
They showed that stellar halos
are  moderately flattened spheroids with spatial luminosity distributions
that are well described by a power law ($\sim r^{-3}$).

On the other hand, there are some disk galaxies with prominent stellar halos
which are bright enough to be visually identified.
A well known example is NGC 4594 (M104), a spiral galaxy, known as
the Sombrero galaxy, with a Hubbe type of Sa \citep{dev91} however, it is
sometimes considered as a peculiar elliptical galaxy \citep{mcc16} because
of its bright stellar halo.
The nucleus of NGC 4594 is fairly bright, and there is a supermassive black
hole in the center \citep{kor88, kor96, ho97}. More often, the prominent
stellar halo of NGC 4594 is considered as a massive bulge which contains
$\sim90\%$ of the total luminosity \citep{ken88}. However, the
$V-$band minor axis profile of NGC 4594 \citep{jar85} shows a break
at $\sim55$ arcsec which suggests two components, a stellar halo and central
bulge. Detailed surface photometry of NGC 4594 \citep{bur86} also showed
that the outer spheroidal component is a prominent stellar halo that can be
well approximated by an ellipsoid with an axis ratio of $b/a=0.64$.
A clear distinction between the central bulge and the stellar halo can be
seen in the Spitzer 3.6-$\mu$ image \citep{gad12}, in which
the stellar halo is rounder than the central bulge, and
dominates over the disk and bulge at large radius ($r \gtrsim 215$ arcsec).

%%%%%%%%%%%%%%%% Fig~1 (Mrur.eps) %%%  ) %%%%%%%%%%%%%%%%%
\begin{figure}[t!]
\centering
\includegraphics[trim=0mm 10mm 0mm 0mm, clip, width=1\columnwidth]{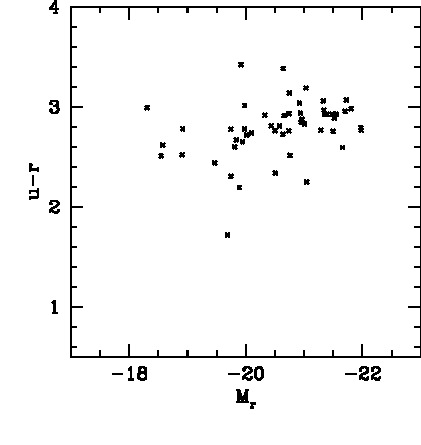}
\caption{Color-magnitude diagram of the sample galaxies that show prominent stellar halos.
\label{fig1}}
\end{figure}
%\end{verbatim}%%%%%%%%%%%%%%%%%%%%%%%%%%%%%%%%%%%%%%%%%%%%%%%%%%%%%%%

Since most stellar halos are too faint to be observed by conventional
methods, the number of galaxies with detailed observations of their
stellar halos is quite limited. Therefore, there are no statistical samples
of the structure and physical properties
of stellar halos. On the other hand, stellar halos have been well
predicted by numerical simulations \citep{kat91, ste95, bro04} in the CDM
scenario \citep{whi91, kau93}.
In a sense, these simulations failed to reproduce Milky Way type
stellar halos whose mass fraction is less than $\sim1$ per cent of the total
luminous matter. Most stellar halos that they modeled were too massive
compared to observed stellar halos. However, \citet{bro04} was able to make
different kinds of stellar halos depending on the supernova (SN)
feedback mechanisms used.
One model is similar to the faint stellar halos, often observed, and another
is similar to the massive stellar halos that were reported in the previous
studies \citep{kat91, ste95}.
Although prominent stellar halos can easily be observed by conventional
methods, there is no extensive study of the prominent stellar halos due to
their rarity in the local universe. Observations of prominent
stellar halos, however, seem to be a promising path to understanding the
structural properties of stellar halos. This study will uncover a wealth of
information about the formation and  evolution of galaxies.

The purpose of this study is to construct a sample of
prominent stellar halos and establish basic statistics.
In particular, we aim to derive the S\'ersic index of the luminosity profile
of the prominent stellar halos to characterize the shape of the luminosity
distribution. The S\'ersic index $n$ is known to be correlated with the luminosity, effective radius, and the
central velocity dispersion of galaxies \citep{gra01, mol01, gra02}
when applied to ellipticals and the bulges of disk galaxies,
Since prominent stellar halos are
rare phenomena, we are also interested in their environment.

This paper is organized as follows. The data and sample selection are
described in Section~\ref{sec:data}, and the results of the present study are given in
Section~\ref{sec:results}. Discussion and conclusions are provided in Section~\ref{sec:discuss}.

%%%%%%%%%%%%%%%% Fig~2 (images.eps) %%%  ) %%%%%%%%%%%%%%%%%
\begin{figure*}[p]
\centering
\includegraphics[width=0.95\textwidth]{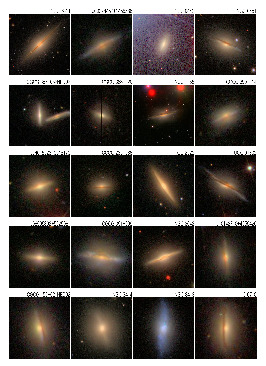}
\caption{Images of disk galaxies that show prominent stellar halos.
The image sizes are adjusted to fit to the frame size.
\label{fig2}}
\end{figure*}
%\end{verbatim}%%%%%%%%%%%%%%%%%%%%%%%%%%%%%%%%%%%%%%%%%%%%%%%%%%%%%%%

%%%%%%%%%%%%%%%%% Fig~2 (images.eps) %%%  ) %%%%%%%%%%%%%%%%%
\addtocounter{figure}{-1}
\begin{figure*}[p]
\centering
\includegraphics[width=0.95\textwidth]{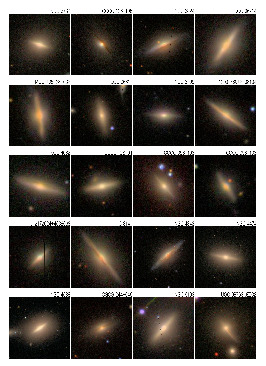}
\caption{\emph{Continued.}}
\end{figure*}
%\end{verbatim}%%%%%%%%%%%%%%%%%%%%%%%%%%%%%%%%%%%%%%%%%%%%%%%%%%%%%%%

%%%%%%%%%%%%%%%%% Fig~2 (images.eps) %%%  ) %%%%%%%%%%%%%%%%%
\addtocounter{figure}{-1}
\begin{figure*}[t!]
\centering
\includegraphics[width=0.95\textwidth]{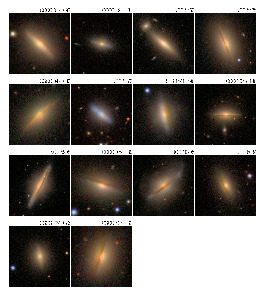}
\caption{\emph{Continued.}}
\vskip3\baselineskip
\end{figure*}
%\end{verbatim}%%%%%%%%%%%%%%%%%%%%%%%%%%%%%%%%%%%%%%%%%%%%%%%%%%%%%%%

\section{Data and Sample Selection \label{sec:data}}

\subsection{Selection of Prominent Stellar Halos}

We selected prominent stellar halos using the color images provided by the
Sloan Digital Sky Survey (SDSS) Data Release 7 \citep{ab09}.
We used the Korea Institute of Advanced Study Value-Added Galaxy
Catalog \citep[KIAS-VAGC;][]{chk10} to select the parent sample of target
galaxies that have redshifts less than $z = 0.032$, and $r$-magnitudes brighter
than 17.77. The number of parent sample galaxies is 44,244. We first selected
7,810 edge-on galaxies through visual inspections of the SDSS color
images in the parent sample. Because of the large axis ratios ($b/a$) of the
prominent stellar halos, they are likely to be omitted in the sample of edge-on
galaxies selected by axis ratio criteria.
Thus, it is necessary to conduct a visual inspection of the color
images to see whether or not galaxies are edge-on. The selection criterion we
have applied is the axis ratio of the disks and not the entire galaxy.
Sometimes, a well distinguished dustlane is considered as evidence of
edge-on disks.

Since we have selected the prominent stellar halos by visual
inspections of the color images without quantitative selection criteria,
the selected sample is subject to some personal bias, but we do not think
we omit a significant fraction of prominent stellar halos. The final sample
contains 54 galaxies, and corresponds
to $0.1\%$ of the parent sample of 44,244 galaxies, and $0.7\%$ of the 7810
edge-on galaxies. This means that prominent stellar halos are very rare.
The basic properties of the 54 galaxies
are presented in Table~\ref{tab1}, most of which are
obtained from the KIAS-VAGC.

Figure \ref{fig1} shows the color-magnitude diagram of galaxies in the final sample.
M$_{r}$ is derived from the $r-$band model magnitude corrected for Galactic
extinction using the $r-$band extinction values from SDSS DR7.
A K-correction and an evolution correction are not applied.
Distances to the galaxies were derived from the redshifts, corrected for
the motion relative to the centroid of the local Group \citep{mou00}.
For galaxies with $z < 0.01$, the metric distances provided by NED are used.
The galaxies lying inside a $10^{\circ}$-cone around M87 with
redshift less than $z = 0.007$ are assumed to be the members
of the Virgo Cluster, and the distance of the Virgo cluster is used
for those galaxies. We assumed the Virgo distance as $D=16.7$ Mpc and
H=75 km s$^{-1}$ Mpc$^{-1}$.
It is evident that the galaxies with
prominent stellar halos have photometric properties similar to the galaxies
in the red sequence defined by early type galaxies, particularly
by elliptical galaxies.

\section{Results \label{sec:results}}

\subsection{Morphological Properties of Prominent Stellar Halos}

Figure \ref{fig2} shows the color images of the 54 sample galaxies from the
SDSS DR7. At first glance, the prominent stellar halos are mostly
flattened ellipsoids.
They have somewhat red colors, typical of old stellar populations, and sizes
comparable to their disks. Some halos have sizes about half that of their
disk, while some halos are nearly spherical. For the majority of sample
galaxies dust lanes can be observed, some of which are offset from
the central bulge, and middle plane of the disk.
One of the remarkable morphological
properties of galaxies with prominent stellar halos is that they are
early type disk galaxies, mostly S0 and S0/a galaxies, without well developed
bulges. Most of the luminosity around the disk come from the
stellar halo, and not from the bulge. The bulge luminosity is confined to the
central regions. Of course, as shown in previous studies \citep{hes93},
the prominent structures around the disk plane
can be considered as an extended bulge. However, as will be shown below,
the extraplanar prominent structures are considered here
as stellar halos.

As can be seen in Figure~\ref{fig2}, most galaxies with prominent stellar
halos have red disks except for the two extremely blue
disks (NGC 3413 and NGC 5672) and show well distinguished dustlanes.
There are several galaxies whose disk morphology differs from that of
others. They have disks with somewhat blue colors in the outer parts,
indicating spiral arms (2MASX J00224457+1456586, UGC04332, CGCG 091-099,
NGC 4343, CGCG 191-031, UGC 10205). Warped disks are observed in a small
fraction of sample galaxies.

\begin{figure}[t!]
\centering
\includegraphics[width=1\columnwidth]{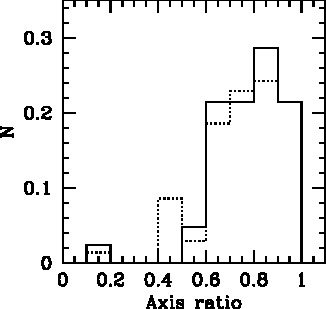}\vspace{2mm}
\caption{Frequency distribution of axis ratios of the disk galaxies which
have prominent stellar halos, compared to elliptical galaxies in
the local universe ($z\lesssim0.01$). Solid lines represent the galaxies with
prominent stellar halos and dotted lines indicate the local elliptical
galaxies.
\label{fig3}}
\end{figure}

Figure \ref{fig3} shows the histogram of the axial ratios ($b/a$) of the sample
galaxies along with the axial ratios of a sample of
elliptical galaxies in the local
universe ($z\lesssim0.01$) from the catalog of \citet{ann15}.
The axis ratios of the sample galaxies were taken from the
KIAS-VAGC \citep{chk10} where the major- and minor-axis length were derived
from the $i$-band isophotal maps.
the two types of galaxies in the local universe show a similar distribution of
axis ratios. Since the lengths of the major axis of the stellar halos
and the disks are similar, the axial ratios of the sample galaxies are
determined by the shape of the halos. Thus, the similarity
between the two axis ratio distributions suggests that the prominent stellar
halos are flattened ellipsoids, similar to the elliptical galaxies.

%%%%%%%%%%%%%%%% Fig~3 (HSHb2a.eps) %%%  ) %%%%%%%%%%%%%%%%%
%\end{verbatim}%%%%%%%%%%%%%%%%%%%%%%%%%%%%%%%%%%%%%%%%%%%%%%%%%%%%%%%

\subsection{Luminosity Profiles of the Prominent Stellar Halos}

Surface photometry of the $r$-band images for the sample galaxies is used to
derive the luminosity profiles of the prominent stellar halos. We used the
mode of pixel values around the target galaxy as the sky background, $I_{sky}$,
to derive the galaxy intensity distribution given by
$I_{g}=I_{obs}(x,y) - I_{sky}$.
We used the SPIRAL \citep{ich87, oka88} to extract the surface brightness
profiles along the minor-axis, where the luminosity of the stellar halo is
less affected by the disk luminosity. The effect of dustlanes is also
negligible in the minor-axis profiles, especially in the halo regions.

The bulge and halo are the main contributors to the luminosity profile along
the minor-axis. In most normal disk galaxies
the contribution of halo component is
negligibly small compared to that of the bulge. But, in the case of
disk galaxies with prominent stellar halos, the luminosity of the stellar
halo is larger than that of the bulge. In particular,
the luminosity distribution at radii larger than a few times the effective
radius of the bulge is dominated by the stellar halo. The surface
brightness profiles along the minor-axis of the sample galaxies are examined
to see  whether the sample galaxies have two components.
Breaks in the surface brightness profiles are considered as the signature of
multiple components. The majority of the sample galaxies are found to
have a break at $r \lesssim 10$ arcsec from the center, indicating the
existence of a central bulge.

Several functional forms are commonly used to represent the luminosity profiles
of galaxies. In the early era of galaxy studies, the Hubble profile,
$I(r)=I_{0}/(1+(r/r_{c})^2)$, was employed to represent the luminosity
distribution of elliptical galaxies. Later, \citet{dev58} introduced the
$r^{1/4}$-law  to represent the luminosity distributions of the bulge of
spiral galaxies as well as the elliptical galaxies. \citet{fre70}
introduced exponential function to describe the luminosity distribution of
the disks of spiral galaxies. These functions have been used very successfully
to describe the luminosity distributions of ellipticals and spirals.
\citet{ser68} introduced a generalized function to represent luminosity
distribution of galaxies including both ellipticals and spirals.
The S\'ersic function
is defined by three parameters, $I_{eff}$, $r_{eff}$, and $n$ as follows
$$I(r)=I_{eff}exp(-b_{n}((r/r_{eff})^{1/n} - 1)$$
where $r_{eff}$ and $I_{eff}$ are the effective radius and effective intensity
(which is the intensity at $r=r_{eff}$), respectively. The parameter $n$ is
the S\'ersic index. The constant $b_{n}$ is approximated
by $b_{n}=1.9992n-0.3271$ \citep{cao93, pru97}. The de Vaucouleurs law and
exponential law are the cases for $n=4$ and $n=1$, respectively.

As there is no commonly used functional form for the luminosity profile
of prominent stellar halos, both the Hubble profile and the S\'ersic profile
are examined to determine which is more representative of the luminosity
distribution. The Hubble profile is known to
represent the luminosity distribution of the stellar halos of
edge-on galaxies obtained by image stacking techniques \citep{zib04},
However, it was found that the Hubble profile does not fit well
to the observed luminosity distributions of the majority of the sample
galaxies in our sample.
On the contrary, the S\'ersic profile better represents the luminosity
distribution of the prominent stellar halos.

Figure \ref{fig4} shows the results of fitting a single S\'ersic profile along the
minor-axis. A $\chi^{2}$ minimization technique is applied to
derive the three free
parameters, $n$, $r_{eff}$ and $I_{eff}$, with fitting ranges selected from the
visual inspection of the minor-axis profiles. The bulge dominated regions
are avoided during fitting a single S\'ersic profile to the observed profiles.
The resulting profiles are insensitive to the fitting ranges except in the
far outer regions where the surface brightness is affected by the
background sky or by other components such as rings.
For some galaxies, there are wiggles or bumps at large radii which
are due to the contamination from nearby objects or in the case of large
bumps due to outer rings or shell structures.

At first glance,
the observed luminosity distributions of prominent stellar halos are
well represented by the S\'ersic profiles. As shown in Figure~\ref{fig5},
the fitted
S\'ersic index $n$ varies from 0.5 to 3.5 with a peak near $n=1.2$.
About $90\%$ of the sample galaxies have $n$ less than 3.
For the majority of galaxies. there is a luminosity
excess over the halo profile at small radii. This
central luminosity excess is mostly due to the bulge which is
not taken into account in the profile fittings. There are 46 galaxies
in the sample($85\%$) that show a clear luminosity excess.
Since the luminosity excess is due to the luminosity of the bulge,
the luminous structures around the disk plane must be stellar halos.
Thus, this study presents a statistically significant sample of the
structures of stellar halos for the first time.
The stellar halos are prominent enough to be visually
identified from conventional images, such as SDSS. The three
parameters of the single component S\'ersic profile are listed in Table~\ref{tab2}.

\begin{table*}
\centering
\caption{Basic properties of 54 disk galaxies with prominent stellar halos.\label{tab1}}
\begin{tabular}{rrrrrrrrrrl}
\toprule
ID  & $\alpha_{2000}$  & $\delta_{2000}$  & D$^{\rm a}$  & M$_{r}$  & $u-r$  & $g-r$  & $mc^{\rm b}$  & $b/a$  & $\sigma^{\rm c}$ & Galaxy name$^{\rm d}$  \\
\midrule
 354822 &   0.812042 &  16.145420 & 12.6 & -19.92 & 3.37 & 1.28 & SA(s)ab? & 0.25 &   0.0 & NGC 7814                   \\
 381062 &   5.685914 &  14.949751 & 77.0 & -19.89 & 2.19 & 0.76 & Sa  & 0.81 &  98.1 & J00224457+1456586         \\
2092266 &  18.661020 &  -1.020657 & 56.4 & -21.34 & 2.97 & 1.04 & S0/a?  & 0.64 & 176.4 & NGC 0442                   \\
 323194 &  27.295120 & -10.426420 & 17.8 & -19.98 & 2.96 & 1.06 & SAB(s)ab & 0.68 & 108.0 & NGC 0681                   \\
1742502 &  32.784565 &  -0.654782 & 60.3 & -20.76 & 2.51 & 0.85 & Sb    & 0.70 & 125.1 & CGCG 387-052NED02          \\
 809401 &  37.040436 &   0.799052 & 74.7 & -21.01 & 2.83 & 0.97 & S0/a  & 0.90 & 193.5 & CGCG 388-020               \\
1767358 &  40.438672 &   0.442579 & 10.2 & -18.31 & 1.64 & 0.88 & SBb?  & 0.52 &  74.5 & NGC 1055                   \\
 764571 &  47.772167 &  -0.546510 & 84.0 & -20.51 & 2.34 & 0.78 & S0/a  & 0.74 & 127.7 & CGCG 390-014                \\
 326159 &  61.230087 &  -6.304820 & 80.1 & -20.96 & 2.87 & 0.99 &    1 & 0.89 &   0.0 & J04045523-0618170         \\
1728488 & 116.389717 &  45.772331 & 94.8 & -21.48 & 2.92 & 1.01 & S0   & 0.83 & 180.6 & CGCG 235-036              \\
1905088 & 121.556000 &  17.706573 & 45.8 & -21.33 & 3.06 & 1.06 & S0/a  & 0.53 & 273.9 & NGC 2522                   \\
1875243 & 124.907845 &  21.114300 & 53.7 & -20.65 & 3.38 & 0.98 & S?  & 0.72 & 176.1 & UGC 04332                  \\
 225814 & 132.721054 &  55.419109 & 91.6 & -20.74 & 2.76 & 0.94 & S0   & 0.73 & 184.5 & J08505302+5525081         \\
2257869 & 142.595779 &  19.469234 & 42.4 & -19.74 & 2.31 & 0.92 &   2 & 0.93 &  71.6 & CGCG 091-099               \\
  78806 & 148.334244 &   0.697720 & 35.7 & -20.93 & 2.94 & 1.03 & S0    & 0.70 & 180.4 & NGC 3042                   \\
1248278 & 153.612778 &  40.945045 & 68.5 & -19.98 & 3.01 & 1.11 & S0/a  & 0.68 & 125.8 & J10142710+4056420         \\
1960169 & 162.291870 &  29.815895 & 94.0 & -21.35 & 2.93 & 1.03 &   1 & 0.63 &   0.0 & CGCG 155-021NED02         \\
2204882 & 162.817932 &  27.975279 & 13.9 & -19.84 & 2.85 & 0.97 & S0    & 0.64 &   0.0 & NGC 3414                   \\
1894107 & 162.836288 &  32.766331 &  6.1 & -16.07 & 2.27 & 0.38 & S0    & 0.57 &   0.0 & NGC 3413                   \\
2221022 & 173.614166 &  25.876469 & 93.3 & -21.39 & 2.92 & 1.04 &   1 & 0.63 & 186.3 & IC 0710                    \\
 965668 & 173.901520 &  54.948639 & 59.0 & -21.50 & 2.76 & 0.94 & SB0   & 0.77 & 202.4 & NGC 3737                   \\
2356878 & 174.083298 &  20.521761 & 63.0 & -20.33 & 2.91 & 1.02 &    1 & 0.72 & 228.3 & CGCG 126-108               \\
 873264 & 175.686890 &  52.779636 & 57.6 & -20.95 & 2.84 & 1.01 & SA(s)a?  & 0.43 & 146.0 & NGC 3824                   \\
 253847 & 176.162704 &  67.955879 & 28.8 & -18.91 & 2.61 & 0.92 & S?  & 0.70 &  74.6 & UGC 06714                  \\
2205307 & 177.191284 &  29.641171 & 68.4 & -20.44 & 2.81 & 1.02 &    2 & 0.59 &   0.0 & MCG +05-28-032             \\
 901271 & 177.644882 &  50.529064 & 69.8 & -21.29 & 2.77 & 0.93 & S0/a  & 0.58 & 203.3 & UGC 06811                  \\
2322850 & 179.485378 &  25.068284 & 41.0 & -18.58 & 2.62 & 0.88 &    1 & 0.69 &  87.1 & NGC 3999                   \\
2333089 & 180.823410 &  20.636995 & 61.5 & -20.10 & 2.74 & 0.95 &    1 & 0.60 & 117.8 & J12031760+2038134         \\
1061991 & 181.297729 &  10.670613 & 68.5 & -20.66 & 2.91 & 1.05 & Sbc   & 0.83 & 146.0 & NGC 4082                   \\
2247528 & 182.475800 &  25.311249 & 68.5 & -20.50 & 2.76 & 0.95 &    1 & 0.80 & 166.9 & CGCG 128-04                \\
2414948 & 183.345688 &  16.957827 & 83.9 & -20.58 & 2.81 & 0.93 &    1 & 0.55 & 185.4 & CGCG 098-103               \\
2311827 & 183.505875 &  23.010292 & 72.7 & -19.75 & 2.78 & 0.96 &    2 & 0.77 & 114.5 & J12140140+2300370         \\
1202224 & 184.334381 &  46.635284 & 71.4 & -20.02 & 2.72 & 0.90 & S0/a  & 0.90 & 187.7 & J12172024+4638068         \\
2319495 & 184.743393 &  24.186310 & 72.8 & -20.75 & 2.93 & 1.03 &    2 & 0.62 & 200.2 & IC 3141                    \\
1288874 & 185.914588 &   6.951942 &  9.1 & -18.92 & 3.00 & 1.02 & SA(rs)b?  & 0.63 & 101.0 & NGC 4343                   \\
1828259 & 187.473083 &  14.068640 & 15.2 & -19.47 & 2.63 & 0.91 & S0   & 0.56 &   0.0 & NGC 4474                   \\
 959348 & 190.697632 &  11.442470 & 10.9 & -19.81 & 2.80 & 0.95 & S0   & 0.53 &   0.0 & NGC 4638                   \\
1176719 & 193.205093 &  47.685013 & 96.5 & -20.92 & 3.04 & 1.05 &    1 & 0.55 &   0.0 & CGCG 244-046               \\
1227719 & 200.125336 &  43.083969 & 13.7 & -18.55 & 2.71 & 0.89 & S?  & 0.86 &   0.0 & NGC 5103                   \\
1989264 & 207.772507 &  25.093710 & 87.5 & -21.97 & 2.79 & 0.93 &    1 & 0.79 &   0.0 & UGC 08763NED02             \\
  49109 & 208.428314 &   0.060874 & 88.4 & -21.52 & 2.89 & 0.99 & S0/a  & 0.86 & 221.2 & CGCG 017-093               \\
1838833 & 210.311371 &  37.882977 & 74.2 & -20.64 & 2.72 & 0.95 & Sa   & 0.55 & 128.5 & CGCG 191-031               \\
1340227 & 211.543945 &   9.353449 & 72.2 & -21.98 & 2.76 & 0.94 & S?   & 0.55 &   0.0 & NGC 5463                   \\
2380372 & 213.913589 &  14.282601 & 55.3 & -21.51 & 2.89 & 0.98 & S0   & 0.67 & 226.5 & NGC 5525                   \\
1348148 & 217.970367 &   7.320885 & 81.3 & -21.03 & 3.19 & 1.12 &    2 & 0.69 & 173.3 & CGCG 047-080               \\
1913169 & 218.159805 &  31.670155 & 36.2 & -19.69 & 1.71 & 0.56 & Sb?  & 0.95 &   0.0 & NGC 5672                   \\
 980767 & 220.880341 &  49.393093 & 91.1 & -21.05 & 2.25 & 0.82 & Sa   & 0.83 & 137.4 & SBS 1441+496               \\
1430278 & 225.568130 &  11.917582 & 97.1 & -21.81 & 2.98 & 1.03 & Sa   & 0.85 & 203.8 & CGCG 076-136               \\
2412381 & 229.180084 &  55.409248 & 34.9 & -20.75 & 3.14 & 1.11 & SA(s)b  & 0.56 & 111.9 & NGC 5908                   \\
1423817 & 241.207611 &   8.481091 & 52.8 & -19.94 & 2.65 & 0.91 &    1 & 0.64 & 120.3 & CGCG 079-039               \\
1393362 & 241.667435 &  30.099058 & 67.1 & -21.66 & 2.59 & 0.95 & Sa   & 0.58 & 165.2 & UGC 10205                  \\
2001255 & 243.569672 &  17.757429 & 91.1 & -21.72 & 3.07 & 1.07 & Sa   & 0.65 & 212.5 & NGC 6084                   \\
 532728 & 249.074478 &  44.135700 & 96.0 & -21.70 & 2.95 & 1.01 & S0/a  & 0.86 & 220.8 & CGC G224-076               \\
 726224 & 328.157928 &  12.535856 & 89.1 & -21.55 & 2.93 & 1.06 & E   & 0.67 & 222.9 & CGCG 427-032               \\
\bottomrule
\end{tabular}
\tabnote{
  $^{\rm a}$ Galaxy distance in unit of Mpc.\\
  $^{\rm b}$ Morphological types form the NASA Extragalactic Database. If not available, the KIAS-VAGC morphological class is given: 1 (E/S0) and 2 (Sp/Irr).\\
  $^{\rm c}$ Central velocity dispersion in unit of km s$^{-1}$. In case
  of no data, we use 0.0.\\
  $^{\rm d}$ Galaxy name taken from the NASA Extragalactic Database.
  We omit `2MASX' for the 2 Micron All Sky Survey Extended objects names.
}
\end{table*}

Since the majority of sample galaxies show a central luminosity excess, which
is assumed to be due to bulge component, the luminosity distributions along
the minor-axis of sample galaxies are also
fitted by two S\'ersic functions, one for
the bulge and the other for the halo. The iterative fitting
technique \citep{kor77} is applied to decompose the bulge and halo
simultaneously.
The iterative profile decomposition technique assumes fitting ranges for each
component where one component dominates the other.
A small change in the bulge fitting
range affects the decomposition significantly. Thus, the bulge fitting range
is selected interactively by minimizing the residuals of the fit.
At first, $n=4$, (i.e., de Vaucouleurs law), was assumed
for the bulge, and $n$ was varied for the halo. Using these parameters
the fittings were unsuccessful for the majority of the sample galaxies.
Various alternative values of $n$ were explored for the bulge, and the
S\'ersic index of $n\approx1$ is found to well represent most galaxies.
Since $n=1$ gives successful fits for most of the sample
galaxies, $n=1$ is used as the default S\'ersic index for the bulge
component. However, $n$=4 was adopted if $n=4$ gives a similar result.
In general, the S\'ersic index of the stellar halo
in the two component model becomes smaller than that of the single component
model. The S\'ersic index of the two component models $n_{b}$ (bulge)
and $n_{h}$ (halo) are listed in Table~\ref{tab2}.

%%%%%%%%%%%%%%%% Fig~4 (serfit.eps) %%%  ) %%%%%%%%%%%%%%%%%
\begin{figure*}[t!]
\centering
\includegraphics[trim=12mm 0mm 0mm 0mm, clip, width=1\textwidth]{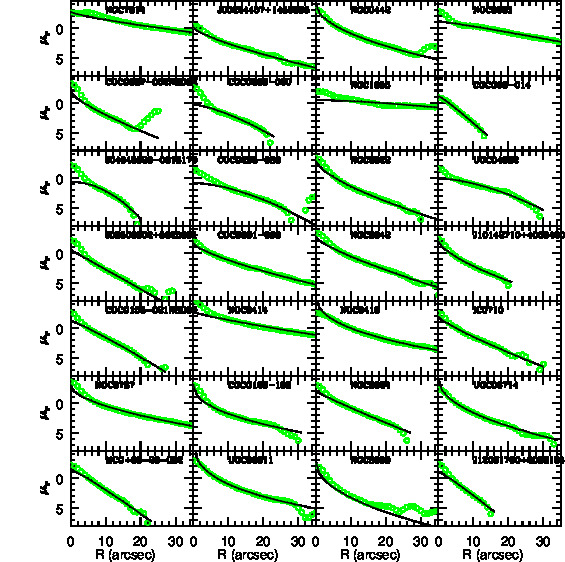}
\caption{Single component S\'ersic function fit to the minor-axis profiles of
galaxies that show prominent stellar halos. The green circles
indicate the data points while the black curve is the fitting result.}
\label{fig4}
\vskip1\baselineskip
\end{figure*}
%\end{verbatim}%%%%%%%%%%%%%%%%%%%%%%%%%%%%%%%%%%%%%%%%%%%%%%%%%%%%%%%

%%%%%%%%%%%%%%%% Fig~4 (serfit.eps) %%%  ) %%%%%%%%%%%%%%%%%
\addtocounter{figure}{-1}
\begin{figure*}[t!]
\centering
\includegraphics[trim=0mm 0mm -5mm -47mm, clip, width=1\textwidth]{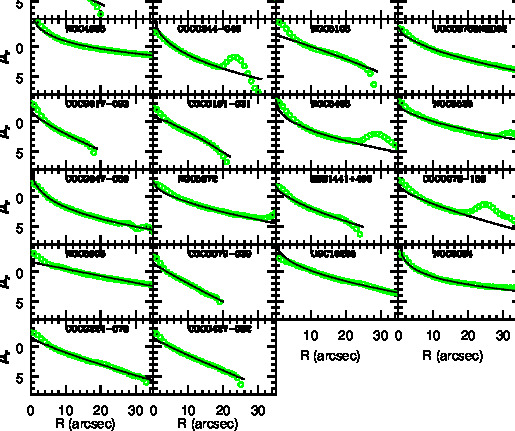}
\caption{\emph{Continued.}}
\vskip1\baselineskip 
\end{figure*}
%\end{verbatim}%%%%%%%%%%%%%%%%%%%%%%%%%%%%%%%%%%%%%%%%%%%%%%%%%%%%%%%

It is interesting to note that about $15\%$ of the sample galaxies
have shell structures outside of the visible stellar halo. The presence of
the shell structure is indicated by the large bump in the minor-axis
profiles. The galaxies with well developed outer shell structure are
NGC 0442, CGCG 387-052NED0, NGC 3999,  NGC 4082, CGCG 098-103, CGCG 244-046,
NGC 5463, 2MASX J14443669+1631314, CGCG 076-136.

%%%%%%%%%%%%%%% Fig~5 (Hnser.eps) %%%  ) %%%%%%%%%%%%%%%%%

\begin{figure}[t!]
\centering
\includegraphics[trim=10mm 10mm 0mm 0mm, clip, width=1\columnwidth]{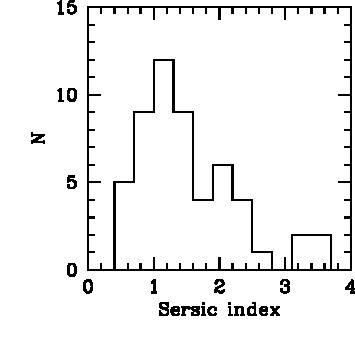}
\caption{Frequency distribution of the S\'ersic index for the single component
model.}
\label{fig5}
\end{figure}

%\end{verbatim}%%%%%%%%%%%%%%%%%%%%%%%%%%%%%%%%%%%%%%%%%%%%%%%%%%%%%%%

\subsection{Distribution of the S\'ersic Index}

Figure \ref{fig5} shows the frequency distribution of the S\'ersic index for the
54 stellar halos derived by fitting a single S\'ersic function
to the surface brightness profiles
along the minor-axis of the sample galaxies. Since the detailed shape of the
frequency distribution depends on the bin size, due to the small sample size,
the frequency distributions using different bin sizes were examined.
It was found that the bin size of  $\Delta n=0.3$ is optimum choice for the
present sample. It keeps the general shape of the frequency distribution,
while suppressing the statistical noise.

It appears that there are at least two peaks, one peak at $n\approx1.2$,
and the other at $n\approx3.4$. The number of galaxies around the first
peak at $n\approx1.2$ is 37 and it comprises $\sim70\%$ of the total sample.
The number of galaxies near the second peak at $n\approx3.4$ is 4, which
is less than $10\%$ of the total sample. The remaining galaxies,
comprising the remaining quarter of the total sample, are located near a
third peak at $n\approx2$. Since it is not clear whether the third peak at
$n\approx2$ is a real feature, the sample galaxies are divided into two groups.
One group is comprised of the galaxies around the first peak
at $n\approx1.2$, including
the galaxies near the third peak at $n\approx2$, and the other group consists
of galaxies around the second peak at $n\approx3.4$.

The number of galaxies in the larger group is 50 which comprises $93\%$ of
the total sample. The mean S\'ersic index of this group is $1.42\pm0.55$. If
we exclude the galaxies near the third peak at $n\approx2$, it becomes
$1.19\pm0.35$, thus very close to the exponential profile. The distribution
of the S\'ersic index of this group is similar to that of the dwarf
elliptical/spheroidal galaxies \citep{bj98, ryd99,grant05, kmk06,
jan14, seo18}.

\begin{figure}[t!]
\centering
\includegraphics[trim=7mm 8mm 0mm 0mm, clip, width=1\columnwidth]{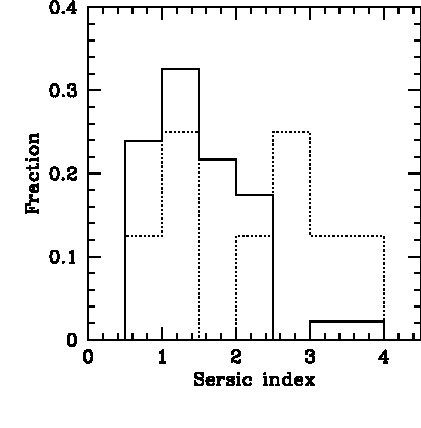}
\caption{Frequency distribution of the S\'ersic index for the single component
model grouped by the presence/absence of a central luminosity excess.
Galaxies with a central excess are represented by the solid line while
those without central excess are plotted by dotted line.
}
\label{fig6}
\end{figure}

Figure \ref{fig6} shows the frequency distribution of the S\'ersic index of
prominent stellar halos with (solid lines) and without (dotted lines) the
central luminosity excess, respectively.
The prominent stellar halos of the sample galaxies
with a central luminosity excess have small $n$, while those
without a central luminosity excess show double peaks, with one peak
at $n\approx1.3$ and the other peak at $n\approx2.8$. There are only two
galaxies (4.3\%) that have $n$ larger than 2.4 among galaxies with a central
luminosity excess, while 63\% of galaxies without central luminosity excess
have $n$ larger than 2.2. Thus, the extraplanar prominent structure
without central luminosity excess can be thought of as the extended bulges.

\subsection{Dependence on the Physical Properties of Galaxies}

\begin{figure}[t!]
\centering
\includegraphics[trim=7mm 7mm 0mm 0mm, clip, width=1\columnwidth]{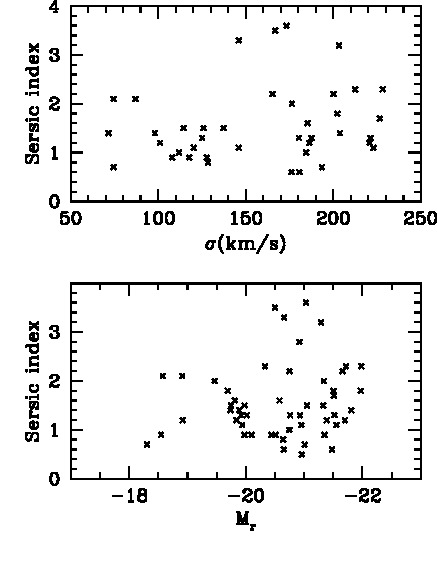}
\caption{Correlations between physical parameters and the S\'ersic index for
the single component model. $n$-$\sigma$ in the upper panel and
$n$-M$_{r}$ in the lower panel.
}
\label{fig7}
\end{figure}

The dependence of the S\'ersic index $n$, derived from the single-component
fitting, on the two observable physical parameters, the $r$-band absolute
magnitude (M$_{r}$), and the velocity dispersion ($\sigma$), are investigated.
Both parameters are closely related to the mass of a galaxy, which is a key
parameter that controls the formation and evolution of a galaxy \citep{pen12}.
Figure~\ref{fig7} shows the distribution of $n$ as a function of
M$_{r}$ (lower panel) and $\sigma$ (upper panel). As shown in the lower panel
of Figure~\ref{fig7}, there is no clear correlation between the S\'ersic index
and the luminosity. There may be, however, a tendency of smaller $n$
for less luminous galaxies (M$_{r} > -20$). More specifically,
there are no prominent stellar halos that have $n$ larger than $\sim2$ for
galaxies with M$_{r} > -20$, but prominent stellar halos with $n$ larger than
$\sim2$ are observed in bright galaxies (M$_{r} < -20$). However, even for
bright galaxies, their stellar halos are likely to have the S\'ersic index
of $n\approx2$.

The upper panel of Figure~\ref{fig7} shows the distribution of $n$ as a
function of $\sigma$.
There is no strong correlation between the two parameters, but
there is a tendency of increasing $n$ with
increasing $\sigma$. It is also apparent that there are  no prominent
stellar halos that have a large S\'ersic index ($n > 2.2$) for galaxies
with velocity dispersions less than $\sigma = 150$ km s$^{-1}$. Thus,
the dependence of the S\'ersic index on the velocity dispersion is very similar
to the dependence of the S\'ersic index on the luminosity.

%%%%%%%%%%%%%%% Fig~6 (Exhn54.eps %%%  ) %%%%%%%%%%%%%%%%%
%\end{verbatim}%%%%%%%%%%%%%%%%%%%%%%%%%%%%%%%%%%%%%%%%%%%%%%%%%%%%%%%

The structure of a galaxy formed through major mergers is different from that
formed through a monolithic collapse of galactic proto-clouds. Giant
ellipsoidal structures (S\'ersic index of $n\approx4$) are likely to be made
by violent mergers, while disks (S\'ersic index of $n\approx1$) are made by
dissipational collapse of proto-clouds that have non-negligible angular
momentum. The absence of large S\'ersic index ($n > 2.2$) for galaxies with
M$_{r} > -20$ or $\sigma < 140$ km $^{-1}$, i.e, relatively low mass galaxies,
suggests that major mergers do not play a significant role for less luminous
galaxies with prominent stellar halos. It is worth noting that our dividing
value of $n=2.2$ is the same as the critical value suggested by \citet{fis08}
who used this value to divide pseudobulges from classical bulges.
If galaxies with small S\'ersic indices did not form through major mergers,
then the majority of  prominent stellar halos must be  caused by other
mechanisms. This is because $\sim70\%$ of prominent stellar halos
have a S\'ersic index less
than the critical value, regardless of luminosity or mass.

%%%%%%%%%%%%%%% Fig~7 (hnMrvsig2.eps) %%%  ) %%%%%%%%%%%%%%%%%
%\end{verbatim}%%%%%%%%%%%%%%%%%%%%%%%%%%%%%%%%%%%%%%%%%%%%%%%%%%%%%%%

\begin{table}[t!]
\centering
\caption{S\'ersic index of prominent stellar halos.\label{tab2}}
\begin{tabular}{rrrrrrr}
\toprule
ID  & $n$ & $u_{eff}^{\rm a}$ & $r_{eff}^{\rm b}$ & $n_{b}$ & $n_{h}$ & $\Sigma_{5}^{\prime}$ $^{\rm c}$ \\
\midrule
 354822 & 1.3 & -0.16 &  1.84 & 1.0 &  0.6 &  -1.470  \\
 381062 & 1.4 &  2.56 &  4.82 & 4.0 &  1.7 &  -1.980  \\
2092266 & 2.0 & -0.07 &  2.39 & 1.0 &  0.8 &   0.142  \\
 323194 & 0.9 &  0.54 &  2.15 & 1.0 &  0.5 &  -0.399  \\
1742502 & 1.3 &  0.74 &  2.35 & 1.0 &  0.3 &  -0.138  \\
 809401 & 0.7 &  1.48 &  3.96 & 1.0 &  0.4 &  -0.428  \\
1767358 & 0.7 &  0.68 &  2.32 & 1.0 &  0.5 &  -0.255  \\
 764571 & 0.9 &  0.46 &  2.19 & 0.9 &  0.9 &  -0.063  \\
 326159 & 0.5 &  1.41 &  3.72 & 1.0 &  0.4 &  -1.670  \\
1728488 & 0.6 &  1.82 &  6.58 & 1.0 &  0.4 &   0.040  \\
1905088 & 1.5 &  0.08 &  1.74 & 1.0 &  0.7 &   0.187  \\
1875243 & 0.6 &  1.08 &  3.88 & 1.0 &  0.4 &   0.718  \\
 225814 & 1.0 &  1.33 &  3.36 & 1.0 &  0.6 &   1.290  \\
2257869 & 1.4 &  1.07 &  2.64 & 4.0 &  1.2 &  -0.425  \\
  78806 & 1.3 & -0.12 &  1.74 & 4.0 &  1.3 &  -1.030  \\
1248278 & 1.5 &  0.93 &  2.62 & 1.0 &  0.9 &  -0.347  \\
1960169 & 0.9 &  0.40 &  3.77 & 1.0 &  0.5 &   0.579  \\
2204882 & 1.2 & -0.40 &  1.48 & 1.0 &  0.5 &  -0.015  \\
1894107 & 2.5 & -0.56 &  0.22 & 1.0 &  0.8 &  -0.143  \\
2221022 & 1.2 &  0.44 &  3.93 & 1.0 &  0.5 &   0.186  \\
 965668 & 1.8 &  0.71 &  4.23 & 4.0 &  0.9 &   1.310  \\
2356878 & 2.3 &  1.54 &  3.68 & 1.0 &  0.5 &   0.627  \\
 873264 & 1.1 &  0.05 &  2.60 & 1.0 &  0.8 &  -0.658  \\
 253847 & 2.1 &  0.14 &  0.86 & 1.0 &  0.3 &   0.007  \\
2205307 & 0.9 &  0.19 &  2.29 & 1.0 &  0.2 &  -0.362  \\
 901271 & 3.2 & -0.86 &  2.03 & 1.0 &  0.2 &   0.790  \\
2322850 & 2.1 &  1.39 &  1.23 & 1.0 &  1.4 &   1.820  \\
2333089 & 0.9 &  0.55 &  1.80 & 1.0 &  0.5 &   0.757  \\
1061991 & 3.3 &  2.92 &  3.92 & 1.0 &  0.3 &  -0.706  \\
2247528 & 3.5 &  1.01 &  2.52 & 1.0 &  0.3 &   0.844  \\
2414948 & 1.6 &  1.22 &  3.50 & 1.0 &  1.0 &  -0.394  \\
2311827 & 1.5 &  1.57 &  3.07 & 1.0 &  0.7 &  -0.167  \\
1202224 & 1.3 &  1.10 &  2.52 & 1.0 &  0.5 &   0.218  \\
2319495 & 2.2 &  2.20 &  2.70 & 0.2 &  1.0 &  -0.001  \\
1288874 & 1.2 & -0.43 &  0.41 & 0.2 &  1.0 &   1.830  \\
1828259 & 2.0 &  0.76 &  1.47 & 4.0 &  1.3 &   1.120  \\
 959348 & 1.6 &  2.06 &  1.49 & 4.0 & 10.0 &   1.060  \\
1176719 & 2.8 & -1.02 &  1.73 & 1.0 &  1.0 &   0.338  \\
1227719 & 0.9 & -0.26 &  0.65 & 1.0 &  0.4 &  -0.409  \\
1989264 & 1.8 &  0.25 &  5.30 & 1.0 &  0.6 &   1.740  \\
  49109 & 1.3 &  0.65 &  3.30 & 1.0 &  0.6 &  -0.052  \\
1838833 & 0.8 &  0.65 &  3.06 & 1.0 &  0.4 &  -0.547  \\
1340227 & 2.3 &  1.01 &  3.93 & 0.2 &  0.4 &   0.710  \\
2380372 & 1.7 &  0.45 &  4.71 & 1.0 &  0.6 &  -0.358  \\
1348148 & 3.6 &  0.14 &  2.66 & 1.0 &  1.0 &   0.100  \\
1913169 & 1.8 &  1.05 &  2.43 & 0.2 &  1.0 &  -0.064  \\
 980767 & 1.5 &  1.17 &  4.17 & 1.0 &  0.4 &  -0.560  \\
1430278 & 1.4 &  0.83 &  5.30 & 0.2 &  1.0 &   0.137  \\
2412381 & 1.0 &  0.24 &  2.66 & 1.0 &  0.4 &   0.450  \\
1423817 & 1.1 &  0.62 &  1.95 & 1.0 &  1.2 &  -0.524  \\
1393362 & 2.2 & -0.51 &  3.71 & 1.0 &  0.4 &  -0.660  \\
2001255 & 2.3 &  3.94 &  4.12 & 0.2 &  1.0 &  -0.630  \\
 532728 & 1.2 &  0.71 &  5.41 & 1.0 &  0.6 &   1.120  \\
 726224 & 1.1 &  0.58 &  3.91 & 1.0 &  0.8 &   1.100  \\
\bottomrule
\end{tabular}
\tabnote{
  $^{\rm a}$ Effective surface brightness - $\mu_{sky}$. \\
  $^{\rm b}$ Effective radius in unit of arcsec. \\
  $^{\rm c}$ $\Sigma_{5}^{\prime}$ is defined as $\log(\Sigma_{5}/\bar{\Sigma_{5}})$.
}
\end{table}

\subsection{Local Background Density}

The local background density ($\Sigma$) is used
to examine the environmental dependence of the S\'ersic index. The
local  background density can be derived using a variety of
approaches \citep{mul12}, but the $n$th nearest neighbor method with $n=5$
is used here. The definition of neighbor galaxy by \citet{ann17} is used
where a neighboring galaxy is defined to have a
velocity difference less than the linking velocity
$\Delta V=500$ km s$^{-1}$, and a luminosity brighter than the
limiting magnitude of the volume-limited sample.
The background density using the projected distance to the $5th$ nearest
galaxy is defined as ,
$$ \Sigma_{5}={5 \over 4\pi r_{p, 5}^{2}},$$
\noindent{where ${r_{p, 5}}$ is the projected distance to the 5th nearest
neighbor galaxy. The local background density is normalized by the mean
background density ($\bar{\Sigma_{5}}$).}

\begin{figure}[t!]
\centering
\includegraphics[trim=5mm 5mm 0mm 0mm, clip, width=1\columnwidth]{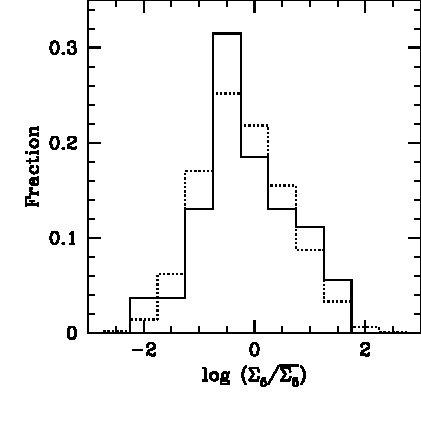}
\caption{Frequency distribution of the local background density of the
prominent stellar halos (solid lines) compared with SDSS galaxies
that have redshift less than $z=0.04$.
}
\label{fig8}
\end{figure}

Figure \ref{fig8} shows the histogram of the local background density of the prominent
stellar halos along with that of SDSS galaxies with redshift less than
$z=0.04$. There is not much difference between the two distributions.
But, the present sample shows somewhat larger fractions at high density regions
and smaller fractions at low density regions than the full sample. If we divide the full sample into the blue galaxies and red galaxies,
the local background density of the present sample is more similar to that
of red galaxies. The low density tail of the present sample is due to
the spiral galaxies.

Since there is a tight relationship between the
luminosity of a galaxy and the local background density \citep{goto03, park07},
we used galaxies brighter than M$_{r}=-20$ to examine the environmental
dependence of the S\'ersic index. Fig. 9 shows the frequency
distribution of the local background density for prominent stellar halos
for galaxies brighter than M$_{r}=-20$. The distribution of prominent
stellar halos with $n < 2.2$ is similar to that of the background
galaxy distribution, while the distribution of the prominent stellar halos with
$n > 2.2$ shows a much narrower distribution, with maximum frequency at
$\Sigma_{5}/\bar{\Sigma_{5}}=0.75$. However, due to the small sample size
for the galaxies with large $n$, this result could be due to
statistical noise. We carried out a K-S test and found that there is no
significant difference between the two groups.
The K-S test gives $p=0.21$.

%%%%%%%%%%%%%%% Fig~8 (Hden54.eps) %%%  ) %%%%%%%%%%%%%%%%%
%%%%%%%%%%%%%%%%%%%%%%%%%%%%%%%%

\section{Discussion and Conclusions \label{sec:discuss}}

We have analyzed the luminosity profiles along the minor-axis of disk galaxies
with luminous halos. Many previous studies
considered the prominent stellar halo to be an extended
bulge, which is well fitted by the $r^{1/4}$-law \citep{wai90, ems96}.
However, in this study it is treated as a stellar halo which is
luminous enough to be
observed. The reason for this is that most of the
sample galaxies have bright central regions that are assumed to be the central
bulge component. They are represented by a
central excess of light in the surface brightness profile along the minor-axis
when the observed profile is fitted by a single S\'ersic profile.
The average S\'ersic index $n$ of the single component fits is $n=1.1\pm0.9$,
and the majority of sample galaxies have $n$ less than $\sim2.2$. The small
$n$ also supports the assumption that the extended regions are
prominent stellar halos rather than
extended bulges. In the literature, there are some
studies that consider the luminous
spheroidal component as a prominent stellar halo \citep{har84, bur86, wag89}
although some authors have not explicitly used the terminology spheroid
\citep{bur86}.

The S\'ersic index $n$ of prominent stellar halos is quite different from
that of classical bulges. Approximately $80\%$ of prominent
stellar halos in the sample have
$n$ smaller than the critical value of $n=2.2$, which divides the bulges of disk
galaxies into classical bulges and pseudobulges \citep{fis08}. Classical bulges
have $n$ greater than 2.2, and pseudobulges have $n$ smaller than 2.2.
Given that the S\'ersic index $n$ is a shape parameter which is closely
related to the galaxy formation mechanism, the majority of the stellar halos
are formed by mechanisms different from that for the classical
bulge formation. Prominent stellar halos do not seem to have formed from
major mergers followed by violent relaxation because resulting substructures
would have large $n$. If the luminous spheroidal components are
bulges, they are likely to be pseudobulges because of their small $n$,
since pseudobulges are thought to be made through the secular evolution of disk
galaxies, driven by non-axisymmetric potential such as
a bar \citep{kor04,oka13}. However, luminous structures
around the disks are not considered to be pseudobulges because of their extent
which is comparable to the disks.
Since pseudobulges are made of disk material, the luminosity of pseudobulges
is not expected to dominate over the disk luminosity.

%%%%%%%%%%%%%%% Fig~9 (MrhnHden54.eps) %%%  ) %%%%%%%%%%%%%%%%%
\begin{figure}[t!]
\centering
\includegraphics[trim=5mm 5mm 0mm 0mm, clip, width=1\columnwidth]{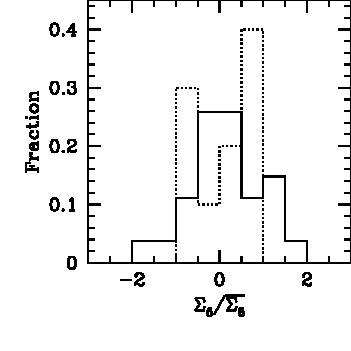}
\caption{Frequency distribution of the local background density of the
prominent stellar halos with $n < 2.2$ (solid lines) compared
with the cases for $n > 2.2$. We used galaxies brighter than
M$_{r}=-20$.
}
\label{fig9}
\end{figure}
%%%%%%%%%%%%%%%%%%%%%%%%%%%%%%%%

We have examined the dependence of the S\'ersic index of prominent stellar
halos on the physical properties of galaxies and their environment.
The physical properties considered here are the galaxy luminosity (M$_{r}$)
and the central velocity dispersion ($\sigma$). As shown in Figure~\ref{fig7},
galaxies with low luminosities (M$_{r} \gtrsim -20$) or low velocity
dispersions ($\sigma \lesssim 150$ km s$^{-1}$) have S\'ersic indices smaller
than $n\approx2.2$, while galaxies with high luminosities and high velocity
dispersions have a wide range of S\'ersic indices including $n > 2.2$.
If a large S\'ersic index is considered to be related to the substructures
formed through merger induced violent relaxation, then the prominent stellar
halos of massive galaxies must have
origins different to those of less massive galaxies.
One plausible scenario for the origin of the  prominent stellar halos
that have $n$ smaller than 2.2 is the intensive $in-situ$ star formation.
The fact that there is no environmental dependency on the prominent
stellar halos supports the above scenario, since supernova explosions
are an internal process which does not depend on the environment.

In the $\Lambda$CDM cosmology, the stellar halos of disk galaxies are formed
by the debris of accreted satellites \citep{bul05, rea06,sal07, fon08,
del08, joh08, fon11} as well as $in-situ$ halo stars \citep{coo15}. Halo stars
formed $in-situ$ include heated disk stars which are ejected into the halo.
These heated disk stars make a negligible contribution to the total halo
mass \citep{coo15}, but they contribute significant fractions in the inner halo,
similar to the fraction of stars from
satellite debris \citep{pur10, fon11, mcc11}. The disrupted debris
of accreted satellites and $in-situ$ halo stars are a good explanation for the
faint stellar halos such as that of the Milky Way or M31. However, they can
not account for the prominent stellar halos presented in this study, because
the luminosity of the prominent stellar halos is brighter than the disk
luminosity.
Since the present sample of prominent stellar halos is observed in early-type
disk galaxies, mostly S0 and S0/a, the cause of prominent stellar halos are
thought to be related to the same feedback mechanism that produces lenticular
galaxies. The thermal feedback model by \citet{bro04} seems to provide a
promising mechanism for producing prominent stellar halos, as well as gas
poor disks such as those of S0 galaxies.
A significant fraction of the stars in
prominent stellar halos can be considered to be due to the $in-situ$ star
formation from the gas ejected from the disk by violent supernova explosions.
It is worth noting that the thermal feedback model in \citet{bro04}
produces stellar halos similar to those shown in the present study, while their
adiabatic feedback model shows stellar halos similar to those of nearby
spiral galaxies such as the Milky Way.

%--------------------------------------------------------------------

\acknowledgments

We would like to thank the anonymous referee for their suggestions and
corrections. We also thank Dong-Kyu Lee for careful reading the manuscript.
This work was supported by the NRF Research grant 2015R1D1A1A09057394.

\end{document}